\documentclass[11pt]{article}

\usepackage[margin=1in]{geometry}
\usepackage{amsmath}
\usepackage{upgreek}
\usepackage{amssymb}
\usepackage{url}

\DeclareMathOperator{\PPp}{{P}_{p}}
\DeclareMathOperator{\Ep}{{E}_{p}}
\DeclareMathOperator{\Varp}{{var}_{p}}
\DeclareMathOperator{\MSEp}{{MSE}_{p}}
\DeclareMathOperator{\mseu}{mse_u}
\DeclareMathOperator{\mseb}{mse_b}
\DeclareMathOperator{\RMSE}{RMSE}
\DeclareMathOperator{\AB}{AB}

\def\leq{\leqslant }
\def\geq{\geqslant }
\newcommand{\pkg}[1]{\texttt{#1}}

\begin{document}

\LARGE
\begin{center}
\textbf{Design-based composite estimation rediscovered}\\[2.00\baselineskip]

\normalsize

\text{Andrius {\v C}iginas}\\[1.50\baselineskip]

{\footnotesize
Vilnius University\\
[2.00\baselineskip]}
\end{center}

\begin{abstract}

\normalsize

Small area estimation methods are used in surveys, where sample sizes are too small to get reliable direct estimates of parameters in some population domains. We consider design-based linear combinations of direct and synthetic estimators and propose a two-step procedure to approach the optimal combination. We construct the mean square error estimator suitable for this and any other linear composition that estimates the optimal one. We apply the theory to two design-based compositions analogous to the empirical best linear unbiased predictors (EBLUPs) based on the basic area- and unit-level models. The simulation study shows that the new methods are efficient compared to estimation using EBLUP. \vskip 2mm 


\end{abstract}

\normalsize

\section{Introduction}\label{s:1}  

Traditional direct estimators can be inefficient if there are additional needs to estimate parameters for unplanned domains of the survey population. The direct estimator is based only on the domain sample that can be too small to get accurate estimates. In the small area estimation theory \cite{RM_2015}, the estimation domain (area) is called small if the direct estimator has there an unacceptably high variance. The theory considers the alternative indirect estimators based on linking models, which help to borrow sample information from neighbor domains through auxiliary data available from registers or other surveys. That approach increases the effective sample size and hence reduces the variances of estimators in the small area. The disadvantage of these estimators is their biases, while the direct estimators are unbiased or approximately so.

Synthetic estimators based on implicit linking models and their linear combinations with the direct estimators constitute an important subclass of the indirect estimators. They are considered in the traditional design-based estimation theory \cite[Chapter 3]{RM_2015}, where estimators of parameters are based only on the randomness induced by the sampling design. The composite estimation is the way to find a trade-off between large variances of the direct estimators and biases of the synthetic estimators. Even some modern indirect estimators, like the empirical best linear unbiased predictors (EBLUPs) \cite{FH_1979, BHF_1988}, built using linear mixed models, are expressed as the linear combinations of the direct and synthetic estimators. Explicit small area models, like those including random area-specific effects for EBLUPs, are a flexible tool in complex estimations. Therefore, these days they are considered superior to the traditional estimators. On the other hand, the latter design-based approach is desirable in sample surveys, and the estimators are pretty simple.  

Despite the simplicity of the traditional design-based compositions themselves, they are less attractive due to the difficulty in estimating their design mean square errors (MSEs) and especially the bias parts of MSEs. We derive a general MSE estimator for any composition that approximates the optimal one. We also propose a new two-step procedure to estimate the optimal linear combination for any pair of direct and synthetic estimators.

We apply our approach to two specific design-based compositions. We combine the weighted sample means with the regression-synthetic estimators based on area-level auxiliary information and the direct generalized regression (GREG) estimators with the indirect GREG estimators in the case of unit-level data. In the simulation experiment, we use data from the Lithuanian Labor Force Survey (LFS) to estimate the proportions of the unemployed and employed in municipalities. We use EBLUPs \cite{FH_1979, BHF_1988}, based on area- and unit-level models, respectively, as benchmarks to judge the quality of the compositions.

\section{Design-based composite estimation}\label{s:2}

\subsection{Preliminary concepts}\label{s:21}

The set ${\cal U}=\{1, \ldots, N \}$ consists of the labels of elements of the finite survey population. The partition ${\cal U}={\cal U}_1\cup\cdots\cup{\cal U}_M$ of the population describes the domains of interest, where ${\cal U}_i\cap {\cal U}_j=\emptyset$ as $i\neq j$, and there are $N_i$ elements in the domain ${\cal U}_i$.  Let $y$ be a study variable with the fixed values $y_1,\ldots, y_N$ assigned to the elements of ${\cal U}$. To estimate the domain parameters $\theta_i$, for instance, the domain means 
\begin{equation}\label{dom_means}
\theta_i=\frac{1}{N_i}\sum_{k\in {\cal U}_i}y_k, \qquad i=1,\ldots, M,
\end{equation}
the sample $s\subset {\cal U}$ of size $n<N$ is drawn according to the sampling design $p(\cdot)$. If the design without replacement was not constructed to ensure the samples $s_i=s\cap {\cal U}_i$ of fixed sizes $n_i$ in the domains, then small $n_i$ can be obtained, and then the accuracy of any direct estimators $\hat\theta_i^{\mathrm d}$ of $\theta_i$ is questionable because of large design variances $\psi_i=\Varp(\hat \theta_i^{\mathrm d})$. Hereafter we use the symbols $\PPp$, $\Ep$, $\Varp$, and $\MSEp$ to denote probability, expectation, variance, and MSE calculated according to $p(\cdot)$, respectively. 

An alternative to the direct estimator $\hat\theta_i^{\mathrm d}$ is the synthetic estimator $\hat\theta_i^{\mathrm S}$, which uses the sample of a larger area through the implicit linking model. A typical model stands on the synthetic assumption that the small domain has the same characteristics as the large area \cite[Chapter 3]{RM_2015}. Similarly, direct estimators $\hat \psi_i^{\mathrm d}$ of $\psi_i$ have large design variances themselves for small sample sizes. Therefore, applying the generalized variance function (GVF) approach \cite{W_2007}, the estimators $\hat \psi_i^{\mathrm d}$ are smoothed, and more stable estimators $\hat \psi_i^{\mathrm s}$ are further used.

\subsection{Approximations to optimal compositions}\label{s:22} 

Since the synthetic estimator $\hat\theta_i^{\mathrm S}$ of $\theta_i$ uses larger sample, its design variance is often smaller compared to that of the direct estimator $\hat\theta_i^{\mathrm d}$. However, a contribution of its design bias to MSE  can be substantial if the synthetic assumption is not realistic. To find a balance between larger variances $\psi_i$ of $\hat\theta_i^{\mathrm d}$ and the biases of $\hat\theta_i^{\mathrm S}$, we consider the linear compositions
\begin{equation}\label{gen_lin_comb}
\tilde\theta_i^{\mathrm C}=\tilde\theta_i^{\mathrm C}(\lambda_i)=\lambda_i\hat\theta_i^{\mathrm d}+(1-\lambda_i)\hat\theta_i^{\mathrm S}, \qquad i=1,\ldots, M,
\end{equation}
with coefficients $0\leq \lambda_i\leq 1$. Minimizing the function $\MSEp(\tilde\theta_i^{\mathrm C}(\lambda_i))$ with respect to $\lambda_i$, the optimal weight for the $i$th domain is the population characteristic
\begin{equation}\label{opt_lambda}
\lambda_i^*=\frac{\MSEp(\hat\theta_i^{\mathrm S})-C_i}{\MSEp(\hat\theta_i^{\mathrm d})+\MSEp (\hat\theta_i^{\mathrm S})-2C_i} \quad \text{with} \quad C_i=\Ep(\hat\theta_i^{\mathrm d}-\theta_i)(\hat\theta_i^{\mathrm S}-\theta_i).
\end{equation} 
Applying the assumption $|C_i| \ll \MSEp(\hat\theta_i^{\mathrm S})$ and knowing that the estimator $\hat\theta_i^{\mathrm d}$ is nearly unbiased, a standard approximation used to optimal parameter \eqref{opt_lambda} is \cite[Section 3.3]{RM_2015}
\begin{equation}\label{opt_lambda_approx}
\lambda_i^*\approx \MSEp(\hat\theta_i^{\mathrm S})/(\psi_i+\MSEp (\hat\theta_i^{\mathrm S})).
\end{equation}
However, further evaluation from sample data is still complicated because of difficulties to estimate $\MSEp(\hat\theta_i^{\mathrm S})$. The best general method known in the literature, which does not require any additional synthetic assumptions, is to use the representation \cite[Section 3.2.5]{RM_2015}
\begin{equation}\label{GW}
\MSEp(\hat\theta_i^{\mathrm S})=\Ep (\hat\theta_i^{\mathrm S}-\hat\theta_i^{\mathrm d})^2-\Varp(\hat\theta_i^{\mathrm S}-\hat\theta_i^{\mathrm d})+\Varp(\hat\theta_i^{\mathrm S}),
\end{equation}
which includes the unbiased direct estimator $\hat\theta_i^{\mathrm d}$, and then to build an approximately design unbiased estimator
\begin{equation}\label{GWest}
\mseu (\hat\theta_i^{\mathrm S})=(\hat\theta_i^{\mathrm S}-\hat\theta_i^{\mathrm d})^2-\hat\sigma^2(\hat\theta_i^{\mathrm S}-\hat\theta_i^{\mathrm d})+\hat\sigma^2(\hat\theta_i^{\mathrm S})
\end{equation}
of \eqref{GW}, where $\hat\sigma^2(\cdot)$ stands for an estimator of the design variance $\Varp(\cdot)$. However, estimator \eqref{GWest} can be very unstable for individual small domains, and thus it is not efficient to use it for estimation of weight \eqref{opt_lambda} or its approximation \eqref{opt_lambda_approx}.

Therefore, non-straightforward ways are used to approximate and estimate the optimal coefficients for compositions \eqref{gen_lin_comb}. One of the ideas is to set a common weight for all domains (or groups of them) and then minimize a total MSE with respect to that weight \cite{PK_1979}. A similar but more sophisticated composite estimation is to apply James--Stein method \cite[Section 3.4]{RM_2015}. A flexible proposal is sample-size-dependent estimation \cite{DSC_1982}, where estimators of the weights $\lambda_i$ in \eqref{gen_lin_comb} are taken to be dependent on the sample sizes in the domains.

\subsection{Estimation of mean square errors}\label{s:23}

Estimation of MSEs of the design-based composite estimators is a difficult task, as pointed several times in \cite[Chapter 3]{RM_2015}. That is due to estimation of the component $\MSEp(\hat\theta_i^{\mathrm S})$, and estimated weights $\hat\lambda_i$ add more complexity. The main problem here is to estimate biases of the estimators, while we can always apply at least resampling methods to evaluate the design variances.

The general method used for the synthetic estimators can be applied to the compositions as well, see \cite[Example 3.3.1]{RM_2015} and \cite{C_2020}. That is, treating the composition $\hat\theta_i^{\mathrm C}=\tilde\theta_i^{\mathrm C}(\hat\lambda_i)$ as a synthetic estimator, one can use the estimator
\begin{equation*}\label{GWest_comp}
\mseu (\hat\theta_i^{\mathrm C})=(\hat\theta_i^{\mathrm C}-\hat\theta_i^{\mathrm d})^2-\hat\sigma^2(\hat\theta_i^{\mathrm C}-\hat\theta_i^{\mathrm d})+\hat\sigma^2(\hat\theta_i^{\mathrm C})
\end{equation*}
of $\MSEp(\hat\theta_i^{\mathrm C})$. However, this estimator has the same drawbacks as \eqref{GWest} including undesirable property to take negative values.

We construct the estimator of MSE for any composite estimator $\hat\theta_i^{\mathrm C}$ defined by \eqref{gen_lin_comb} that is close to the optimal combination $\hat\theta_i^{\mathrm {opt}}=\tilde\theta_i^{\mathrm C}(\lambda_i^*)$. First consider general composition \eqref{gen_lin_comb} with a fixed weight. Assuming that its direct component $\hat\theta_i^{\mathrm d}$ is nearly unbiased, we have
\begin{equation}\label{pseudo_bias}
\widetilde D_i=\Ep(\tilde\theta_i^{\mathrm C})-\theta_i \approx (1-\lambda_i) B_i, \quad \text{where} \quad B_i=\Ep(\hat\theta_i^{\mathrm S})-\theta_i
\end{equation}
denotes the design bias of the synthetic part. Assuming additionally that $\max \{|C_i|, \Varp (\hat\theta_i^{\mathrm S})\} \ll \psi_i$, optimal parameter \eqref{opt_lambda} is approximated by the quantity $\tilde\lambda_i^*=B_i^2/(\psi_i+B_i^2)$. Assume next that the number $\lambda_i$ in \eqref{gen_lin_comb} is chosen so that it is close to the optimal $\lambda_i^*$. Then, solving the approximate equation $\tilde\lambda_i^*\approx\lambda_i$, we obtain $B_i^2\approx\lambda_i\psi_i/(1-\lambda_i)$. Inserting the latter relation into the square of \eqref{pseudo_bias}, we arrive to
\begin{equation}\label{gen_bias}
\widetilde D_i^2\approx \lambda_i(1-\lambda_i)\psi_i.
\end{equation}
For any design-based composite estimator $\hat\theta_i^{\mathrm C}\approx \hat\theta_i^{\mathrm {opt}}$, we derive the squared estimator $\widehat D_i^2$ of the bias $D_i=\Ep(\hat\theta_i^{\mathrm C})-\theta_i$ by letting $D_i^2\approx \widetilde D_i^2$ and then replacing the unknown parameters in \eqref{gen_bias} by their empirical analogs. Finally, we get the estimators
\begin{equation}\label{mse_est_new}
\mseb (\hat\theta_i^{\mathrm C})=\hat \lambda_i(1-\hat \lambda_i)\hat\psi_i^{\mathrm s}+\hat\sigma^2(\hat\theta_i^{\mathrm C}), \qquad i=1,\ldots, M,
\end{equation} 
of $\MSEp(\hat\theta_i^{\mathrm C})$, where the term $\hat\sigma^2(\hat\theta_i^{\mathrm C})$ is an estimator of the design variance $\Varp (\hat\theta_i^{\mathrm C})$.

\subsection{Procedure of composite estimation}\label{s:24} 

We propose a straightforward procedure to estimate optimal weight \eqref{opt_lambda} through approximation \eqref{opt_lambda_approx} and employing MSE estimation by \eqref{mse_est_new}. In the first step, we ignore the bias $B_i$ of the synthetic estimator in \eqref{opt_lambda_approx} and take the estimator 
\begin{equation*}
\hat\lambda_i^{(1)}=\hat\sigma^2(\hat\theta_i^{\mathrm S})/(\hat\psi_i^{\mathrm s}+\hat\sigma^2(\hat\theta_i^{\mathrm S}))
\end{equation*}
of \eqref{opt_lambda}, and then $\hat m_i^{(1)}=\mseb (\tilde\theta_i^{\mathrm C}(\hat\lambda_i^{(1)}))$ is the MSE estimator of the respective composition. However, if the squared bias $B_i^2$ is not negligible, the naive weight $\hat\lambda_i^{(1)}$ is very likely to underestimate the optimal coefficient $\lambda_i^*$. Therefore, in the second step, we treat the initial composition 
\begin{equation}\label{init_comp}
\tilde\theta_i^{\mathrm C}(\hat\lambda_i^{(1)})=\hat\lambda_i^{(1)}\hat\theta_i^{\mathrm d}+(1-\hat\lambda_i^{(1)})\hat\theta_i^{\mathrm S}
\end{equation}
as the synthetic estimator and build the new composition
\begin{equation}\label{2step_est}
\hat\theta_i^{\mathrm {Cb}}=\hat\lambda_i^{(2)}\hat\theta_i^{\mathrm d}+(1-\hat\lambda_i^{(2)})\tilde\theta_i^{\mathrm C}(\hat\lambda_i^{(1)}), \quad \text{where} \quad \hat\lambda_i^{(2)}=\hat m_i^{(1)}/(\hat\psi_i^{\mathrm s}+\hat m_i^{(1)}),
\end{equation}
and $\mseb (\hat\theta_i^{\mathrm {Cb}})=\hat\lambda_i^{(2)}(1-\hat\lambda_i^{(2)})\hat\psi_i^{\mathrm s}+\hat\sigma^2(\hat\theta_i^{\mathrm {Cb}})$ is the estimator of $\MSEp(\hat\theta_i^{\mathrm {Cb}})$ according to \eqref{mse_est_new}.

\section{Applications}\label{s:3}

\subsection{Area-level auxiliary data}\label{s:31} 

Assume that the auxiliary data are available as the vector of characteristics $\mathbf{z}_i=(z_{1i}, z_{2i},\ldots, z_{Pi})'$ for the $i$th domain. Denote by $\pi_k=\PPp\{ k\in s\}>0$ the inclusion into the sample probabilities. To estimate parameters \eqref{dom_means}, we combine the weighted sample means
\begin{equation}\label{dir_ratio}
\hat \theta_i^{\mathrm d}=\frac{1}{\widehat N_i}\sum_{k\in s_i}\frac{y_k}{\pi_k}, \quad \text{where} \quad \widehat N_i=\sum_{k\in s_i}\frac{1}{\pi_k}, \qquad i=1,\ldots, M,
\end{equation}
that are approximately unbiased, and the regression-synthetic estimators
\begin{equation}\label{Rsyn}
\hat\theta_i^{\mathrm S}=\mathbf{z}_i' \boldsymbol{\hat\upbeta}, \quad \text{where} \quad \boldsymbol{\hat\upbeta}=\Biggl( \sum_{i=1}^M \frac{\mathbf{z}_i\mathbf{z}_i'}{\hat \psi_i^{\mathrm s}}\Biggr)^{-1} \sum_{i=1}^M \frac{\mathbf{z}_i \hat\theta_i^{\mathrm d}}{\hat \psi_i^{\mathrm s}}, \qquad i=1,\ldots, M,
\end{equation}
which are derived from the basic area-level model for EBLUP ignoring area-specific random effects \cite[Section 4.2]{RM_2015}. Here the quantities $\hat \psi_i^{\mathrm s}$ smooth the direct estimators \cite[p. 185]{SSW_1992}
\begin{equation}\label{dir_ratio_var_est}
\hat \psi_i^{\mathrm d}=\frac{1}{\widehat N_i^2} \sum_{k\in s_i} \sum_{l\in s_i} (1-\pi_k\pi_l/\pi_{kl}) \frac{(y_k-\hat \theta_i^{\mathrm d})(y_l-\hat \theta_i^{\mathrm d})}{\pi_k\pi_l}, \qquad i=1,\ldots, M,
\end{equation}
of the design variances $\psi_i$, where $\pi_{kl}=\PPp\{ k, l\in s\}>0$ is the probability that both of the elements $k$ and $l$ will be included into the sample. If the design $p(\cdot)$ is complex, the assumption $\pi_{kl}\approx\pi_k\pi_l$, $k\neq l$, is often used and then variances \eqref{dir_ratio_var_est} are approximated by
\begin{equation}\label{var_approx_a}
\hat \psi_i^{\mathrm d}\approx\frac{1}{\widehat N_i^2} \sum_{k\in s_i} \frac{1}{\pi_k}\left(\frac{1}{\pi_k}-1\right) (y_k-\hat \theta_{i}^{\mathrm d})^2, \qquad i=1,\ldots, M,
\end{equation}
The choice of the GVF method to construct the synthetic variances $\hat \psi_i^{\mathrm s}$ depends on the data. We specify the smoothing in Section \ref{s:4}.

\subsection{Unit-level auxiliary data}\label{s:32} 

Let $\mathbf{x}_k=(1, x_{2k},\ldots, x_{Pk})'$ be the vector containing the values of auxiliary variables $x_2,\ldots, x_P$ for the $k$th element of the population ${\cal U}$. Assume that the data $\mathbf{x}_k$ are available for $k\in s$, and the vector $\boldsymbol{\uptheta}_{xi}=\sum_{k\in {\cal U}_i} \mathbf{x}_k/N_i$ of means is known for the $i$th area. To estimate the domain means, we combine the approximately unbiased direct GREG estimators \cite{SSW_1992, RM_2015}
\begin{equation}\label{dir_GR}
\hat \theta_i^{\mathrm d}=\boldsymbol{\uptheta}'_{xi} \mathbf{\widehat B}_i, \quad \text{where} \quad \mathbf{\widehat B}_i=\Biggl( \sum_{k\in s_i} \frac{\mathbf{x}_k\mathbf{x}'_k}{\pi_k}\Biggr)^{-1} \sum_{k\in s_i} \frac{\mathbf{x}_k y_k}{\pi_k}, \qquad i=1,\ldots, M,
\end{equation}
and their synthetic versions
\begin{equation}\label{GR_syn}
\hat\theta_i^{\mathrm S}=\boldsymbol{\uptheta}'_{xi} \mathbf{\widehat B}, \quad \text{where} \quad \mathbf{\widehat B}=\Biggl( \sum_{k\in s} \frac{\mathbf{x}_k\mathbf{x}'_k}{\pi_k}\Biggr)^{-1} \sum_{k\in s} \frac{\mathbf{x}_k y_k}{\pi_k}, \qquad i=1,\ldots, M,
\end{equation}
called GREG-synthetic estimators \cite[Section 3.2.3]{RM_2015} or indirect GREG estimators. The direct estimators of the design variances $\psi_i$ are \cite{RM_2015}
\begin{equation}\label{dir_GR_var_est}
\hat \psi_i^{\mathrm d}=\frac{1}{\widehat N_i^2} \sum_{k\in s_i} \sum_{l\in s_i} (1-\pi_k\pi_l/\pi_{kl}) \frac{(y_k-\mathbf{x}'_k \mathbf{\widehat B}_i)(y_l-\mathbf{x}'_l \mathbf{\widehat B}_i)}{\pi_k\pi_l}, \qquad i=1,\ldots, M,
\end{equation}
with the simplification
\begin{equation}\label{var_approx_u}
\hat \psi_i^{\mathrm d}\approx\frac{1}{\widehat N_i^2} \sum_{k\in s_i} \frac{1}{\pi_k}\left(\frac{1}{\pi_k}-1\right) (y_k-\mathbf{x}'_k \mathbf{\widehat B}_i)^2, \qquad i=1,\ldots, M,
\end{equation}
used for complex sampling designs.

Weighted sample means \eqref{dir_ratio} are the separate case of direct GREG estimators \eqref{dir_GR}, if we take $P=1$. The components \eqref{dir_GR} and \eqref{GR_syn} of composition \eqref{gen_lin_comb} are similar to that of EBLUP based on the basic unit-level model \cite{BHF_1988}.

\section{Empirical study}\label{s:4}

\subsection{Simulation framework}\label{s:41}

We estimate the proportions of the unemployed and employed in the municipalities of Lithuania. We apply our results for the domain means for each of these two binary study variables, $y$. To draw repeated samples, we create the artificial population from the LFS sample data of the fourth quarter of 2018. To do this, we first remove half of the municipalities due to too small fractions of the unemployed observed in the original sample. Second, we replicate the data of each individual the number of times equal to the rounded survey weight. We get the population ${\cal U}$ of size $N=1396763$ covered by $M=30$ municipalities. We draw $R=10^3$ independent samples of households of size $n'=3700$ without replacement with probabilities proportional to the household sizes. The selected households are surveyed entirely, and then the average size of samples of persons is $n\approx 7667$. We take $\pi_k=h_ln'/N$ for the individual $k\in {\cal U}$, which belongs to the $l$th household of size $h_l$.

We use these binary administrative and demographic variables derived for the estimation quarter: $x_2$ is the indicator that the person is registered as unemployed, $x_3$ marks individuals who paid a social contribution, $x_4$ is to indicate males, and $x_5$ and $x_6$ show the belonging to age groups 26--40 and 41--55, respectively. Using the values $\mathbf{x}_k=(1, x_{2k},\ldots, x_{6k})'$, we take the means $\mathbf{z}_i=\boldsymbol{\uptheta}_{xi}$ for estimators based only on area-level auxiliary data.

Due to small sampling fractions in the municipalities, approximations \eqref{var_approx_a} and \eqref{var_approx_u} to respective direct estimators \eqref{dir_ratio_var_est} and \eqref{dir_GR_var_est} are even more resonable. To smooth these approximations to get $\hat \psi_i^{\mathrm s}$, the GVF method from \cite{D_1995} is suitable for the domain proportions. Its idea is to assume the relation $\psi_i \approx KN_i^\gamma$ and then estimate the parameters $K>0$ and $\gamma\in \mathbb R$ through a log-log regression model. We estimate the design variances of all synthetic and composite estimators using the rescaling bootstrap from \cite{RWY_1992}.

Let $\hat\mu_i^{(r)}$, $r=1,\ldots, R$, be the realizations of any estimator $\hat\mu_i$ of the parameter $\mu_i$, where $\mu_i$ is a proportion or MSE. We evaluate the root mean squared errors (RMSEs) and absolute biases (ABs) of $\hat\mu_i$ using the formulas
\begin{equation}\label{acc_char}
\RMSE(\hat\mu_i)=\Biggl(\frac{1}{R}\sum_{r=1}^R (\hat\mu_i^{(r)}-\mu_i)^2\Biggr)^{1/2} \quad \text{and} \quad \AB(\hat\mu_i)=\Biggl|\frac{1}{R}\sum_{r=1}^R \hat\mu_i^{(r)}-\mu_i\Biggr|, \qquad i=1,\ldots, M.
\end{equation}
To simplify the comparison of estimators, we group the municipalities by the average domain sample size $\bar n_i\approx\Ep (n_i)$ into three classes of equal size and calculate the average of RMSEs and ABs over domains of each class. We get that the $i$th domain is small if $\bar n_i<116$, is medium for $116\leq\bar n_i<159$, and large as $\bar n_i\geq 159$. We use the averages of \eqref{acc_char} over all municipalities as the main measure of accuracy. We divide the average RMSEs and ABs of the design-based estimators by the respective averages calculated for EBLUP. If such a ratio is greater (lower) than one, we say that the estimator is worse (better) than EBLUP.

\subsection{Results for area-level models}\label{s:42}

Assuming that only area-level auxiliary information is available, we compare the weighted sample means $\hat \theta_i^{\mathrm d}$ from \eqref{dir_ratio}, the regression-synthetic estimators $\hat\theta_i^{\mathrm S}$ by \eqref{Rsyn}, their initial compositions \eqref{init_comp}, and the compositions $\hat\theta_i^{\mathrm {Cb}}$ from \eqref{2step_est} with EBLUPs based on the Fay--Herriot model \cite{FH_1979}. The later predictors are evaluated using the function \pkg{mseFH} from \pkg{R} package \pkg{sae} \cite{MM_2015}, where we use the smoothed variances $\hat \psi_i^{\mathrm s}$ and estimate the variance of the random area effects choosing the method of moments. The same function evaluates the MSE estimators for EBLUPs, and we compare them with MSE estimators \eqref{mse_est_new} applied to design-based compositions \eqref{init_comp} and \eqref{2step_est}. The results for the proportions of the unemployed and employed are presented in Tables \ref{tab:1} and \ref{tab:2}, respectively.

\begin{table}[!h]
\caption{The unemployed proportions are estimated using the area-level model. Average RMSEs and ABs of estimators over domain size classes are taken relative to the accuracy for EBLUP.}\label{tab:1}
\centering
\tabcolsep=9pt 
\vspace{1mm} 
\begin{tabular}{lp{10mm}p{10mm}p{10mm}p{10mm}p{10mm}p{10mm}p{10mm}p{10mm}}
\hline\noalign{\smallskip}
 &                \multicolumn{ 4}{c}{Relative average RMSE} &                  \multicolumn{ 4}{c}{Relative average AB} \\
\cline{2-9}\noalign{\smallskip}
Estimator          & \multicolumn{ 4}{c}{Domain size class} & \multicolumn{ 4}{c}{Domain size class} \\

           &        any &         small &      medium &        large &         any &         small &      medium &         large \\
\hline\hline\noalign{\smallskip}

$\hat \theta_i^{\mathrm d}$ &      1.379 &      1.433 &      1.415 &      1.166 &      0.114 &      0.087 &      0.161 &      0.184 \\

$\hat\theta_i^{\mathrm S}$ &      1.019 &      1.089 &      0.910 &      1.024 &      1.808 &      1.597 &      1.888 &      3.257 \\

$\tilde\theta_i^{\mathrm C}(\hat\lambda_i^{(1)})$ &      0.973 &      1.041 &      0.874 &      0.966 &      1.575 &      1.426 &      1.658 &      2.521 \\

$\hat\theta_i^{\mathrm {Cb}}$ &      0.979 &      1.038 &      0.898 &      0.963 &      1.271 &      1.181 &      1.340 &      1.780 \\

$\mseb (\tilde\theta_i^{\mathrm C}(\hat\lambda_i^{(1)}))$ &      1.442 &      1.487 &      1.315 &      1.480 &      1.699 &      1.694 &      1.667 &      1.806 \\

$\mseb (\hat\theta_i^{\mathrm {Cb}})$ &      1.199 &      1.275 &      1.033 &      1.120 &      1.224 &      1.242 &      1.188 &      1.111 \\

\noalign{\smallskip}\hline\noalign{\smallskip}
\end{tabular}  
\end{table}

\begin{table}[!h]
\caption{The employed proportions are estimated using the area-level model. Average RMSEs and ABs of estimators over domain size classes are taken relative to the accuracy for EBLUP.}\label{tab:2}
\centering
\tabcolsep=9pt 
\vspace{1mm} 
\begin{tabular}{lp{10mm}p{10mm}p{10mm}p{10mm}p{10mm}p{10mm}p{10mm}p{10mm}}
\hline\noalign{\smallskip}
 &                \multicolumn{ 4}{c}{Relative average RMSE} &                  \multicolumn{ 4}{c}{Relative average AB} \\
\cline{2-9}\noalign{\smallskip}
Estimator          & \multicolumn{ 4}{c}{Domain size class} & \multicolumn{ 4}{c}{Domain size class} \\

           &        any &         small &      medium &        large &         any &         small &      medium &         large \\
\hline\hline\noalign{\smallskip}

$\hat \theta_i^{\mathrm d}$ &      1.440 &      1.488 &      1.463 &      1.295 &      0.134 &      0.135 &      0.136 &      0.121 \\

$\hat\theta_i^{\mathrm S}$ &      1.032 &      1.068 &      1.014 &      0.976 &      1.526 &      1.420 &      1.496 &      2.101 \\

$\tilde\theta_i^{\mathrm C}(\hat\lambda_i^{(1)})$ &      0.962 &      0.990 &      0.958 &      0.905 &      1.298 &      1.217 &      1.324 &      1.609 \\

$\hat\theta_i^{\mathrm {Cb}}$ &      0.946 &      0.960 &      0.941 &      0.922 &      1.016 &      0.958 &      1.081 &      1.112 \\

$\mseb (\tilde\theta_i^{\mathrm C}(\hat\lambda_i^{(1)}))$ &      1.052 &      1.082 &      1.153 &      0.587 &      1.326 &      1.256 &      1.645 &      0.831 \\

$\mseb (\hat\theta_i^{\mathrm {Cb}})$ &      0.662 &      0.663 &      0.752 &      0.387 &      0.742 &      0.671 &      1.013 &      0.399 \\

\noalign{\smallskip}\hline\noalign{\smallskip}
\end{tabular}  
\end{table}

In the case of unemployment proportions, the average RMSEs of the direct estimator are the largest. The regression-synthetic estimator works much better according to RMSE, but its design biases are large compared to that of EBLUP. The design-based composition $\hat\theta_i^{\mathrm {Cb}}$ corrects for these biases more than the initial composition. The design-based MSE estimator is also better for the former composition, but the MSE estimation for EBLUP is more accurate.

The results for the proportions of the employed are different in that the composition $\hat\theta_i^{\mathrm {Cb}}$ improves EBLUP and preserves similar biases. Moreover, the MSE estimation for this composite estimator is more efficient than the MSE estimator for EBLUP.

\subsection{Results for unit-level models}\label{s:43}

We compare the direct GREG estimators $\hat \theta_i^{\mathrm d}$ given in \eqref{dir_GR}, the GREG-synthetic estimators $\hat\theta_i^{\mathrm S}$ from \eqref{GR_syn}, their first-step compositions \eqref{init_comp}, and the composite estimators $\hat\theta_i^{\mathrm {Cb}}$ by \eqref{2step_est} with EBLUPs based on the basic unit-level model \cite{BHF_1988}. We calculate these EBLUPs applying the function \pkg{eblupBHF} from \pkg{R} package \pkg{sae} \cite{MM_2015}. To get parametric bootstrap MSE estimates for the predictors, we use the function \pkg{pbmseBHF} from the same package. We compare that MSE estimation with MSE estimators \eqref{mse_est_new} applied to compositions \eqref{init_comp} and \eqref{2step_est}. The results for the proportions of the unemployed and employed are in Tables \ref{tab:3} and \ref{tab:4}, respectively.

\begin{table}[!h]
\caption{The unemployed proportions are estimated using the unit-level model. Average RMSEs and ABs of estimators over domain size classes are taken relative to the accuracy for EBLUP.}\label{tab:3}
\centering
\tabcolsep=9pt 
\vspace{1mm} 
\begin{tabular}{lp{10mm}p{10mm}p{10mm}p{10mm}p{10mm}p{10mm}p{10mm}p{10mm}}
\hline\noalign{\smallskip}
 &                \multicolumn{ 4}{c}{Relative average RMSE} &                  \multicolumn{ 4}{c}{Relative average AB} \\
\cline{2-9}\noalign{\smallskip}
Estimator          & \multicolumn{ 4}{c}{Domain size class} & \multicolumn{ 4}{c}{Domain size class} \\

           &        any &         small &      medium &        large &         any &         small &      medium &         large \\
\hline\hline\noalign{\smallskip}

$\hat \theta_i^{\mathrm d}$ &      1.282 &      1.393 &      1.331 &      0.951 &      0.112 &      0.084 &      0.273 &      0.041 \\

$\hat\theta_i^{\mathrm S}$ &      1.135 &      1.350 &      1.051 &      0.761 &      2.081 &      2.223 &      2.929 &      1.092 \\

$\tilde\theta_i^{\mathrm C}(\hat\lambda_i^{(1)})$ &      1.099 &      1.317 &      1.018 &      0.714 &      2.007 &      2.164 &      2.828 &      1.006 \\

$\hat\theta_i^{\mathrm {Cb}}$ &      1.046 &      1.259 &      0.964 &      0.671 &      1.879 &      2.056 &      2.645 &      0.878 \\

$\mseb (\tilde\theta_i^{\mathrm C}(\hat\lambda_i^{(1)}))$ &      2.290 &      2.705 &      2.488 &      0.730 &      2.659 &      2.798 &      3.889 &      1.000 \\

$\mseb (\hat\theta_i^{\mathrm {Cb}})$ &      2.029 &      2.439 &      2.142 &      0.569 &      2.356 &      2.521 &      3.358 &      0.778 \\

\noalign{\smallskip}\hline\noalign{\smallskip}
\end{tabular}  
\end{table}

\begin{table}[!h]
\caption{The employed proportions are estimated using the unit-level model. Average RMSEs and ABs of estimators over domain size classes are taken relative to the accuracy for EBLUP.}\label{tab:4}
\centering
\tabcolsep=9pt 
\vspace{1mm} 
\begin{tabular}{lp{10mm}p{10mm}p{10mm}p{10mm}p{10mm}p{10mm}p{10mm}p{10mm}}
\hline\noalign{\smallskip}
 &                \multicolumn{ 4}{c}{Relative average RMSE} &                  \multicolumn{ 4}{c}{Relative average AB} \\
\cline{2-9}\noalign{\smallskip}
Estimator          & \multicolumn{ 4}{c}{Domain size class} & \multicolumn{ 4}{c}{Domain size class} \\

           &        any &         small &      medium &        large &         any &         small &      medium &         large \\
\hline\hline\noalign{\smallskip}

$\hat \theta_i^{\mathrm d}$ &      1.151 &      1.300 &      1.059 &      1.048 &      0.123 &      0.124 &      0.132 &      0.103 \\

$\hat\theta_i^{\mathrm S}$ &      1.235 &      1.754 &      0.871 &      0.962 &      2.088 &      3.358 &      1.266 &      1.745 \\

$\tilde\theta_i^{\mathrm C}(\hat\lambda_i^{(1)})$ &      1.183 &      1.676 &      0.843 &      0.915 &      1.910 &      3.007 &      1.187 &      1.644 \\

$\hat\theta_i^{\mathrm {Cb}}$ &      0.977 &      1.212 &      0.816 &      0.848 &      1.182 &      1.420 &      0.883 &      1.461 \\

$\mseb (\tilde\theta_i^{\mathrm C}(\hat\lambda_i^{(1)}))$ &      2.347 &      4.035 &      1.337 &      0.864 &      2.453 &      4.025 &      1.368 &      1.134 \\

$\mseb (\hat\theta_i^{\mathrm {Cb}})$ &      1.237 &      1.797 &      0.944 &      0.672 &      0.864 &      0.927 &      0.794 &      0.874 \\

\noalign{\smallskip}\hline\noalign{\smallskip}
\end{tabular}  
\end{table}

For the proportions of the unemployed, the GREG estimator has the largest design RMSEs, and the GREG-synthetic estimator suffers from large biases. The accuracy of the composition $\hat\theta_i^{\mathrm {Cb}}$ is similar to that of EBLUP in terms of the average RMSEs, but the bias correction is smaller than in the area-level case. The MSE estimator for EBLUP is more efficient than the design-based MSE estimator for $\hat\theta_i^{\mathrm {Cb}}$, except for the group of large domains.

The GREG-synthetic estimators of the proportions of the employed have relatively large biases, but the two-step composition $\hat\theta_i^{\mathrm {Cb}}$ significantly reduces them. The errors of this composite estimator and EBLUP are similar. For the medium and large domains, the MSE estimation for the former estimator is more accurate.

\section{Conclusions}\label{s:5}

The direct and traditional synthetic estimators should always be combined in small domains. There are classical ways to estimate the optimal design-based combination, but we construct another two-step composite estimator that is competitive for EBLUPs in the empirical study.

The proposed design-based MSE estimation can be applied to this and any other composition. That estimation is based on the assumption that the composition is close to the optimal one. Therefore, lower accuracy of the MSE estimator is expected in the case of larger deviations from the optimality. We see this happening to the first-step composition. We also get in the experiment that MSE estimation for the second-step composite estimator is accurate if we compare it with the results for EBLUPs. 

The optimality assumption used for the MSE estimation and the two-step composition help avoid straightforward estimation of the bias of the synthetic estimator. The derived MSE estimator is simple and always non-negative. The proposed composition is an adaptive estimator adapting to that MSE estimation.

\bibliographystyle{NAplain}
\bibliography{Ciginas}

\end{document}